\documentclass[fleqn]{2017SCGE}
\usepackage{multirow}
\usepackage{amsmath}
\makeatletter
\def\cpartlineleft#1{\@cpartlineleft#1\@nil}
\def\@cpartlineleft#1,#2\@nil{%
  \omit
  \@multicnt#1%
  \advance\@multispan\m@ne
  \ifnum\@multicnt=\@ne\@firstofone{&\omit}\fi
  \@multicnt#1%
  \advance\@multicnt-#1%
  \advance\@multispan\@ne
  \kern#2
        \leaders\hrule\@height\arrayrulewidth\hfill
        \leaders\hrule\@height\arrayrulewidth\hfill
  \cr
  \noalign{\vskip-\arrayrulewidth}}
\makeatother

\begin{document}

\ensubject{subject}

%%%%%%%%%%%%%%%%%%%%%%%%%%%%%%%%%%%%%%%%%%%%%%%%%%%%%%%
%%% Authors do not modify the information below
%Letter to the Editor
\ArticleType{Article}
%\SpecialTopic{SPECIAL TOPIC: }
\Year{2017}
\Month{January}
\Vol{60}
\No{1}
\DOI{10.1007/s11432-016-0037-0}
\ArtNo{000000}
\ReceiveDate{January 11, 2017}
\AcceptDate{April 6, 2017}
%\OnlineDate{January 1, 2016}
%%%%%%%%%%%%%%%%%%%%%%%%%%%%%%%%%%%%%%%%%%%%%%%%%%%%%%%

%%%  \title{title}{title for citation}
\title{Millimeter Gap Contrast as a Probe for Turbulence Level in Protoplanetary Disks}{Millimeter Gap Contrast as a Probe for Turbulence Level in Protoplanetary Disks}

%%% Corresponding author: ???????
%%%   \author[number]{Full name}{{email@xxx.com}}
%%% General author: ???????
%%%   \author[number]{Full name}{}
 
\author[1,2]{Yao Liu}{{yliu@pmo.ac.cn}}
\author[3]{Gesa H.-M. Bertrang}{}
\author[3]{Mario Flock}{}
\author[4,5]{Giovanni P. Rosotti}{}
\author[1]{Ewine F. van Dishoeck}{}
\author[6]{\\ Yann Boehler}{}
\author[7]{Stefano Facchini}{}
\author[8]{Can Cui}{}
\author[9]{Sebastian Wolf}{}
\author[1]{Min Fang}{}

%%% Author information for page head. 
\AuthorMark{Liu, Y.}

%%% Authors for citation. 
\AuthorCitation{Liu, Y.; Bertrang, G.H.-M.; Flock, M.; Rosotti, G.P.; van Dishoeck, E.F.; Boehler, Y.; Facchini, S.; Cui, C.; Wolf, S.; Fang, M.}

%%% Address.
%%%   \address[number]{Address, City {\rm Postcode}, Country}
\address[1]{Max-Planck-Institut f\"ur Extraterrestrische Physik, Giessenbachstrasse 1, 85748 Garching, Germany}
\address[2]{Purple Mountain Observatory \& Key Laboratory for Radio Astronomy, Chinese Academy of Sciences, Nanjing 210023, China}
\address[3]{Max-Planck-Institut f\"ur Astronomie, K\"onigstuhl 17, D-69117 Heidelberg, Germany}
\address[4]{Leiden Observatory, Leiden University, P.O. Box 9531, NL-2300 RA Leiden, the Netherlands}
\address[5]{School of Physics and Astronomy, University of Leicester, Leicester LE1 7RH, UK}
\address[6]{Univ. Grenoble Alpes, CNRS, IPAG, F-38000 Grenoble, France}
\address[7]{European Southern Observatory, Karl-Schwarzschild-Str. 2, 85748 Garching, Germany}
\address[8]{DAMTP, University of Cambridge, CMS, Wilberforce Road, Cambridge CB3 0WA, UK}
\address[9]{Institut f\"ur Theoretische Physik und Astrophysik, Christian-Albrechts-Universit\"at zu kiel, Leibnizstr. 15, 24118 Kiel, Germany}
%\contributions{}

%%% Abstract.
\abstract{Turbulent motions are believed to regulate angular momentum transport and influence dust evolution in protoplanetary disks. Measuring the strength of 
turbulence is challenging through gas line observations because of the requirement for high spatial and spectral resolution data, and an exquisite determination 
of the temperature. In this work, taking the well-known HD\,163296 disk as an example, we investigated the contrast of gaps identified in high angular resolution 
continuum images as a probe for the level of turbulence. With self-consistent radiative transfer models, we simultaneously analyzed the radial brightness 
profiles along the disk major and minor axes, and the azimuthal brightness profiles of the B67 and B100 rings. By fitting all the gap contrasts measured from 
these profiles, we constrained the gas-to-dust scale height ratio $\Lambda$ to be $3.0_{-0.8}^{+0.3}$, $1.2_{-0.1}^{+0.1}$ and ${\ge}\,6.5$ for the 
D48, B67 and B100 regions, respectively. The varying gas-to-dust scale height ratios indicate that the degree of dust settling changes with radius. 
The inferred values for $\Lambda$ translate into a turbulence level of $\alpha_{\rm turb}\,{<}\,3\times10^{-3}$ in the D48 and B100 regions, which is consistent 
with previous upper limits set by gas line observations. However, turbulent motions in the B67 ring are strong with $\alpha_{\rm turb}\,{\sim}1.2\,{\times}\,10^{-2}$. 
Due to the degeneracy between $\Lambda$ and the depth of dust surface density drops, the turbulence strength in the D86 gap region is not constrained.}

\keywords{protoplanetary disks, radiative transfer, planet formation}

\PACS{97.82.Jw, 95.30.Jx, 97.82.Fs}

\maketitle

%\tableofcontents

%%%%%%%%%%%%%%%%%%%%%%%%%%%%%%%%%%%%%%%%%%%%%%%%%%%%%%%
%%% The main text. 
%\twocolumn\onecolumn
%%%%%%%%%%%%%%%%%%%%%%%%%%%%%%%%%%%%%%%%%%%%%%%%%%%%%%%
\begin{multicols}{2}

\section{Introduction}
\label{sec:intro}

Protoplanetary disks, as the birthplace of planetary systems, always exhibit turbulent motions \cite{Lesur2022}. There are several mechanisms currently discussed as main 
contributors: hydrodynamical instabilities as the vertical shear instability \cite{Goldreich1967, Fricke1968, Flock2017}, convective overstability \cite{Lyra2014, Klahr2014}, 
Zombie vortex stability \cite{Marcus2015, Marcus2016}, and magneto-hydrodynamical instabilities like the magnetorotational instability \cite{Balbus1991, Balbus1996, Balbus1998}. 
Turbulence regulates the angular momentum transport to sustain gas accretion onto the central star \cite{Shakura73a,Pringle1981}, influences the evolution of dust grains 
in disks \cite{Birnstiel2016}, and plays an important role in controlling the dynamics of embedded planets \cite{Kley2012}. Hence, a detailed understanding of disk evolution 
and planet formation requires knowledge of the strength of turbulent motions.

Placing constraints on the turbulence level is also important in interpreting observational data with numerical simulations. In recent years, 
high-resolution images at infrared and (sub-)millimeter wavelengths have shown that gaps and rings are frequently observed in planet-forming 
disks \cite{Avenhaus2018, Long2018, Andrews2018}. These interesting substructures are often thought to be created by 
planet-disk interaction \cite{Dipierro2018,Zhang2018,Liu2019}. The description of the underlying physics relies heavily 
on (magneto-)hydrodynamical simulations in which turbulence strongly affects the resulting depth and number of
gaps \cite{Pinilla2012a,Flock2015,Rosotti2016,Bertrang2018,Dong2018}. As a consequence, the inferred properties 
(e.g., mass and location) and number of the ``unseen'' (proto)planets are dependent on the input strength of turbulence in the simulation.

However, measuring turbulence with gas line observations is very challenging because on the one hand it demands for data at high spatial 
and spectral resolution, and on the other hand thermal motion usually dominates the broadening of lines, leading to substantial difficulties 
when separating its contribution from the measured total line width \cite{Teague2016}. Therefore, the measurement of turbulence via gas 
line data so far is limited to a small number of disks, revealing low turbulent velocities typically below $5\%\,{\sim}\,10\%$
of the local sound speed ($c_s$) \cite{Guilloteau2012,Flaherty2015,Flaherty2017,Teague2018a,Flaherty2020}. An exception is for the DM Tau disk, 
where the measured turbulent velocity approaches $0.25\,{\sim}\,0.33\,c_s$ \cite{Flaherty2020}.

Turbulence also affects the motion of the dust, either in the radial direction or in the vertical one. Dullemond \& Penzlin \cite{Dullemond2018b} suggested 
that the dependence of turbulence on the dust-to-gas mass ratio together with the radial drift of dust particles could be the origin of the ring structures 
commonly found in protoplanetary disks.
%a feedback loop with the turbulence in gas.  The dust perturbations originate from grains drifting (radially) towards pressure maxima, %and therefore, even small
%dust populations can trigger this feedback loop, leading to an increase in the local pressure maxima which then enable trapping of large grains. The observable 
%results of this feedback loop are rings (associated with gaps) that are expected to be confined in the radial direction \cite{Pinilla2012b}. 
By comparing the width of the millimeter continuum emission ring with the pressure scale height of the disk, Dullemond et al. \cite{Dullemond2018} found strong 
evidence of dust trapping operating in all the rings analyzed in their sample, and put constraints on the quantity $\alpha_{\rm turb}/{\rm St}$, where $\alpha_{\rm turb}$ 
is the turbulence parameter, and $\rm{St}$ is the Stokes number of the dust particles. Vertical stirring induced by turbulent motions acts as a counter process 
against the settling of dust grains. Theoretically speaking, millimeter continuum emission is dominated by millimeter-sized dust particles that are located near 
the midplane of the disk. However, material residing in the adjacent rings, located above the midplane, would hide the gap due to beam smearing. How severe 
this smoothening effect is depends on the scale height of millimeter-sized dust grains \cite{Pinte2016}. In stronger turbulent disks, dust grains are more 
vertically distributed, leading to a more substantial reduction on the gap depth. 

Recently, Doi \& Kataoka \cite{Doi2021} discussed the feasibility of analyzing the intensity variation as a function of azimuth on the rings to estimate the 
degree of dust settling. When the disk is optically thin and viewed at an oblique inclination, the optical depth $\tau$ along the line of sight on  
the major and minor axes differs from each other. Such a difference in $\tau$ forms a peak and dip in the brightness profile at the azimuthal angle of 
the major and minor axis, respectively. The ratio between the brightness peak and dip depends on the millimeter dust scale height. 
The authors fit the azimuthal brightness profiles of two rings in the HD\,163296 disk, and constrained the gas-to-dust scale height ratio and therefore the 
turbulence level. In their analysis, the disk is assumed to be vertically isothermal with a fixed midplane temperature profile. How such a simplification 
affects the result, particularly for rings with a large millimeter dust scale height (i.e., high turbulence regions) needs to be investigated.  

In this work, we take the HD\,163296 disk as an example to investigate in detail the link between millimeter 
gap contrasts and the strength of turbulence, and highlight some features and degeneracies that can be encountered. Sect.~\ref{sec:obs} gives an introduction 
about the HD\,163296 disk. The modeling assumptions are presented in Sect.~\ref{sec:modeling}, while the process of dedicated fitting to the 
ALMA image is described in Sect.~\ref{sec:fitalma}. We discuss our results in Sect.~\ref{sec:discussion}. The paper ends up with a summary in Sect.~\ref{sec:summary}.

\begin{figure*}[!t]
 \centering
 \includegraphics[width=0.85\textwidth]{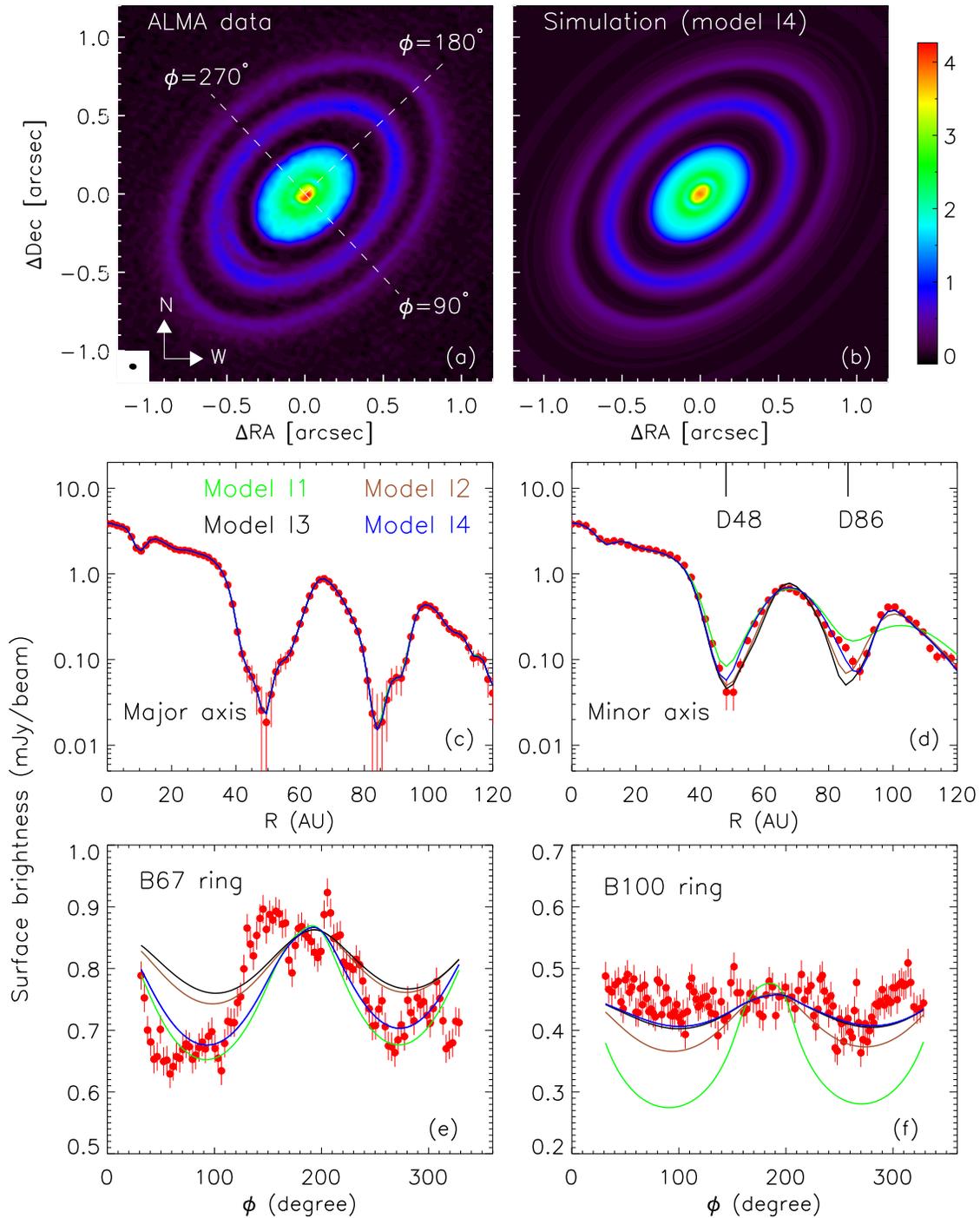}
 \caption{A comparison between models and the DSHARP observation of the HD\,163296 disk. Panel (a): fiducial image generated by the DSHARP team. The beam is 
          shown as the black ellipse in the bottom left corner. The dashed lines indicate the semi-major (to the northwest) and minor axes of the disk. The values 
		  of the azimuth ($\phi$) for the disk major and minor axes are given for a reference of the coordinate. Panel (b): the simulated image of the best-fit 
		  model (i.e., model \texttt{I4}). Panel (c)-(f): a comparison of brightness profiles between observation and different models. The red dots refer to data 
		  points, whereas the green, brown, black and blue lines represent model \texttt{I1}, \texttt{I2}, \texttt{I3} and \texttt{I4}, respectively. Model parameters 
		  can be found in Table~\ref{tab:paras}, and model gap contrasts are compared with the observation in Table~\ref{tab:gapcont}. Note that the four models 
		  overlap well with each other in panel (c).}
\label{fig:imgres}
\end{figure*}

\section{Circumstellar disk of HD\,163296}
\label{sec:obs}
HD\,163296 is a Herbig Ae star (A1 spectral type) located at a distance of $D\,{=}\,101\,{\pm}\,2\,\rm{pc}$ \cite{gaia2018}. 
Its mass ($M_{\star}$) and age are $1.9\,M_{\odot}$ and $10.4\,\rm{Myr}$ \cite{Setterholm2018}. It has a luminosity of $L_{\star}\,{=}\,17\,L_{\odot}$, 
and an effective temperature of $T_{\rm eff}\,{=}\,9250\,\rm{K}$ \cite{Fairlamb2015}. Spatially resolved observations at both infrared and 
millimeter regimes have revealed ring structures in the disk around HD\,163296 \cite{Grady2000,Wisniewski2008,Muro-Arena2018,Isella2016,Notsu2019}. 
Analysis of the interferometric data taken with the Very Large Telescope Interferometer PIONIER and MATISSE yielded brightness 
asymmetries in the near-infrared emission, which may originate from a vortex near the inner rim ($R\,{\sim}\,0.4\,\rm{AU}$) of the disk \cite{Lazareff2017,Varga2021}. 

As one of the 20 targets selected in the Disk Substructures at High Angular Resolution Program (DSHARP), HD\,163296 was 
observed with the Atacama Large Millimeter/submillimeter Array (ALMA) in Band 6 at an unprecedented spatial resolution 
of $4.8\,{\times}\,3.8\,\rm{AU}$ \cite{Andrews2018}. The rms noise of the fiducial ALMA image generated by the 
DSHARP team is $\sigma_{\rm rms}\,{=}\,23\,\mu{\rm Jy/beam}$. The continuum image shows a few pairs of concentric 
rings/gaps, see panel (a) in Figure~\ref{fig:imgres}. The D48 and D86 gaps are located at a radial distance of 48 
and 86\,AU, with a width of 20 and 16\,AU, respectively. The B67 and B100 rings are centered at a radial distance 
of 67 and 100\,AU, with a width of 16 and 12\,AU, respectively \cite{Huang2018}.

We extracted the surface brightness along the disk major and minor axes, given the position angle (PA) of $133.33^{\circ}$. 
Along a PA of $99^{\circ}$, there is a crescent-like structure centered at a radial distance of $55\,\rm{AU}$ \cite{Isella2018}, 
which is probably caused by a Jupiter mass planet \cite{Rodenkirch2021}. Such an asymmetry contaminates the measurement of the 
gap contrast. Hence, we only considered the data on the semi-major axis to the northwest. On the minor axis, however, an 
average of both sides of the disk was performed to improve the signal-to-noise ratio. To apply the methodology introduced 
by Doi \& Kataoka \cite{Doi2021}, we also extracted the azimuthal brightness profiles on the B67 and B100 rings. The reference  
for the azimuthal coordinate ($\phi$) is given in panel (a) of Figure~\ref{fig:imgres}. The extracted brightnesses are shown with 
red dots in panels (c)-(f) of Figure~\ref{fig:imgres}. It should be noted that the mechanism responsible for generating the 
crescent-like structure also likely causes azimuthal perturbations to the B67 ring, which may be one of the reasons why the 
brightness profile shows non-axisymmetric features. The width between two adjacent points (i.e., 1.5\,AU) is about one third 
of the ALMA beam, which means that the brightness is first averaged over such a bin size and then extracted. The errors for each of 
the data points on the major axis, B67 and B100 rings are all set to $23\,\mu{\rm Jy/beam}$, but on the minor axis they 
are calculated to be $\frac{23}{\sqrt{2}}\,\mu{\rm Jy/beam}$ due to the average of both sides of the disk.

\begin{table*}[!t]
\caption{Gap contrasts of the HD\,163296 disk.}
\centering
\footnotesize
\linespread{1.2}\selectfont
\begin{tabular}{lcccccccc}
\hline
\multirow{2}{*}{} & \multicolumn{2}{c}{Major axis} & \multicolumn{2}{c}{Minor axis}  & \multicolumn{2}{c}{B67 ring}  & \multicolumn{2}{c}{B100 ring} \\
                                    \cline {2-3}      \cpartlineleft{4,1em}\cline {5-5}         \cpartlineleft{6,1em}\cline {7-7}      \cpartlineleft{8,1em}\cline {9-9}
                   &         D48                &    D86             &     D48             &    D86         &    $\phi=90^{\circ}$     &    $\phi=270^{\circ}$  &    $\phi=90^{\circ}$     &    $\phi=270^{\circ}$ \\				       		       
\hline
ALMA data &  $0.98\,{\pm}\,0.03$  &  $0.96\,{\pm}\,0.05$  &  $0.94\,{\pm}\,0.02$  &  $0.82\,{\pm}\,0.04$  & $0.22\,{\pm}\,0.03$  &  $0.21\,{\pm}\,0.03$  & $0.00\,{\pm}\,0.07$  &  $0.15\,{\pm}\,0.07$   \\
Model \texttt{I1}   &   0.97    &  0.97   &  0.88   &  0.34   &   0.24   &   0.21  &  0.42  &  0.41 \\   
Model \texttt{I2}   &   0.97    &  0.97   &  0.94   &  0.80   &   0.13   &   0.11  &  0.20  &  0.18 \\  
Model \texttt{I3}   &   0.97    &  0.97   &  0.94   &  0.87   &   0.11   &   0.10  &  0.11  &  0.11 \\  
Model \texttt{I4}   &   0.97    &  0.97   &  0.92   &  0.81   &   0.21   &   0.18  &  0.10  &  0.10 \\
\hline
\end{tabular}
\linespread{1.0}\selectfont
\label{tab:gapcont}
\end{table*}

The gap contrast is defined as $1\,{-}\,I_{\rm min}/I_{\rm max}$, where $I_{\rm min}$ is the minimum brightness within 
the gap, and $I_{\rm max}$ is the maximum brightness of its immediately exterior ring. The brightness profile of the B67 ring displays 
two dips at $\phi\,{=}\,90^{\circ}$ and $270^{\circ}$, which resemble gaps. For simplicity of 
description, we also call them as ``gaps'' hereafter in this work. The constrasts are defined 
as $1\,{-}\,I_{\phi{=}90^{\circ}}/I_{\phi{=}180^{\circ}}$ and $1\,{-}\,I_{\phi{=}270^{\circ}}/I_{\phi{=}180^{\circ}}$. 
On the B100 ring, the profile is quite flat in the western side, and shows only one ``gap'' at $\phi\,{=}\,270^{\circ}$. 
In addition to the chi-square ($\chi^2$) metrics, the observed gap contrasts summarized in Table~\ref{tab:gapcont} are the key characteristics 
used to evaluate the quality of fit of our models. The difference between gap contrasts measured along the disk major and minor axes is 
due to projection effect. Because the disk is geometrically thick, and it is tilted to an inclination of $46.7^{\circ}$, the width 
of the gap varies with azimuthal angle, and reaches the smallest along the minor axis, leading to the lowest gap contrast. 

\section{Full radiative transfer modeling}
\label{sec:modeling}

The key of our work is to constrain the scale height of the millimeter-sized dust grains by fitting the contrasts of gaps with 
self-consistent radiative transfer models, and then link the scale height to the strength of turbulence. In fact, the HD\,163296 disk 
has more gaps, i.e., D10 and D145. However, they are either not fully spatially resolved, or show evidence for being multiple gaps \cite{Huang2018}. 
We will not discuss them in detail throughout the paper, although our modeling methodology automatically captures both features. 

The radiative transfer models are parameterized in the framework of the \texttt{RADMC-3D} code\footnote{http://www.ita.uni-heidelberg.de/~dullemond/software/radmc-3d/.}
\cite{radmc3d2012}. We assume that the disk is passively heated by stellar irradiation. The stellar spectrum is taken from the \texttt{Kurucz} 
database \cite{Kurucz1994}, assuming a gravity of ${\rm log}\,g\,{=}\,3.5$ and solar metallicity. Other model assumptions 
are for the density distribution and dust opacities, which are described below.

\subsection{Dust density distribution}
\label{sec:moddens}

We consider a disk that extends from an inner to outer radii of $R_{\rm in}\,{=}\,0.4\,\rm{AU}$ and $R_{\rm out}\,{=}\,169\,\rm{AU}$, respectively \cite{Huang2018}. 
The model has two distinct dust grain populations, i.e., a small grain population (SGP) and a large grain population (LGP). The temperature structure of the disk 
is mainly governed by the SGP, whereas the LGP dominates the millimeter continuum emission. We fixed the mass fraction of the LGP to $f_{\rm SGP}\,{=}\,0.85$ that has 
been commonly used in previous modeling works of protoplanetary disks \cite{andrews2011,Liu2022}. The SGP is assumed to be well-mixed with the underlying gas distribution. 
Therefore, its scale height is set to the gas scale height ($H_{\rm gas}$) that is solved under the condition of vertical hydrostatic equilibrium. Large dust grains are expected 
to settle towards the midplane \cite{Dubrulle1995,Dullemond2004}. We characterize the degree of dust settling with the parameter $\Lambda$, and the scale height of the 
LGP is given by $H_{\rm gas}/{\Lambda}$.

The volume density of the dust grains is parameterized as 
\begin{equation}
%\rho_{\rm{SGP}}(R,z)\,{=}\,\frac{(1-f_{\rm LGP})\,{\times}\,\Sigma_{\rm d}(R)}{\sqrt{2\pi}\,H_{\rm gas}}\,\exp\left[-\frac{1}{2}\left(\frac{z}{H_{\rm gas}}\right)^2\right], \\
\rho_{\rm{SGP}}(R,z)\,{=}\,\frac{(1-f_{\rm LGP})\,\Sigma_{\rm d}(R)}{\sqrt{2\pi}\,H_{\rm gas}}\,\exp\left[-\frac{1}{2}\left(\frac{z}{H_{\rm gas}}\right)^2\right], \\
\label{eqn:sgp}
\end{equation}
\begin{equation}
%\rho_{\rm{LGP}}(R,z)\,{=}\,\frac{f_{\rm LGP}\,{\times}\,\Sigma_{\rm d}(R)}{\sqrt{2\pi}\,{\times}\,H_{\rm gas}/{\Lambda}}\,\exp\left[-\frac{1}{2}\left(\frac{z}{H_{\rm gas}/{\Lambda}}\right)^2\right], \\
\rho_{\rm{LGP}}(R,z)\,{=}\,\frac{f_{\rm LGP}\,\Sigma_{\rm d}(R)}{\sqrt{2\pi}\,H_{\rm gas}/{\Lambda}}\,\exp\left[-\frac{1}{2}\left(\frac{z}{H_{\rm gas}/{\Lambda}}\right)^2\right], \\
\label{eqn:lgp}
\end{equation}
where $\Sigma_{\rm d}(R)$ is the dust surface density, and $R$ is the distance from the central star measured in the disk midplane. 
Literatural studies usually took analytic forms for $\Sigma_{d}(R)$, e.g., a power law or power law with an exponential taper. However, 
such simple expressions have been demonstrated to be insufficient to capture the fine-scaled features revealed by high resolution 
ALMA observations \cite{Pinte2016,Liu2017}. Instead, we build the surface density by iteratively fitting the 
surface brightnesses at the ALMA wavelength where the optical depth is generally low, see Sect.~\ref{sec:surdens}.

\subsection{Dust properties}
\label{sec:dustopac}

For the dust composition, we made use of the recipe by the DiscAnalysis (\texttt{DIANA}) project \cite{Woitke2016}.
The dust grains consist of 60\% silicate ($\rm{Mg_{0.7}Fe_{0.3}SiO_{3}}$) \cite{dorschner1995}, 15\% amorphous carbon (BE$-$sample) \cite{Zubko1996}, and 
25\% porosity. These percentages are volume fractions of each component, which are used to derive the effective refractory indices of the dust ensemble by 
applying the Bruggeman mixing rule \cite{Bruggeman1935}. We used a distribution of hollow spheres with a maximum hollow volume ratio of 0.8 \cite{Min2005}. 
The mean solid density of the dust ensemble $\rho_{\rm grain}\,{=}\,2.1\,\rm{g\,cm}^{-3}$ is estimated from an average between the silicate 
density ($3.01\,\rm{g\,cm}^{-3}$) and carbon density ($1.8\,\rm{g\,cm}^{-3}$) taking the volume fractions as the weighting factors.

The distribution of grain sizes ($a$) follows a power law ${\rm d}n(a)\,{\propto}\,{a^{-3.5}} {\rm d}a$ with a minimum ($a_{\rm{min}}$) and maximum 
size ($a_{\rm{max}}$). For the SGP, $a_{\rm{min}}$ and $a_{\rm{max}}$ are fixed to $0.01\,\mu{\rm m}$ and $2\,\mu{\rm m}$, respectively. 
For the LGP, $a_{\rm{min}}$ is set to $2\,\mu{\rm m}$. Regarding $a_{\rm{max}}$, we will set it based on models that can reproduce the observed 
millimeter spectral slope, see Sect.~\ref{sec:sedmodel}.

\subsection{Building the dust surface density}
\label{sec:surdens}

Previous studies have shown that surface density profiles in simple analytic expressions (e.g., a smooth power law with density drops at the gap locations)
have difficulties to capture the detailed features revealed by ALMA \cite{Liu2017,Muro-Arena2018}. Using an iterative procedure, we built the surface 
densities by reproducing the millimeter surface brightnesses along the disk major axis that features the maximum spatial resolution. This approach was 
introduced by Pinte et al. \cite{Pinte2016}, and several works by other teams demonstrated its success \cite{Muro-Arena2018,Liu2019}. The iterative 
process consists of the following steps.

\begin{itemize}
\item[a)] We took a starting surface density profile $\Sigma_{\rm d}(R)\,{=}\,\Sigma_{0}\left(R/R_{\rm c}\right)^{-\gamma}{\rm exp}\left[-(R/R_{\rm c})^{2-\gamma}\right]$
with $R_{\rm{c}}\,{=}\,90\,\rm{AU}$ and $\gamma\,{=}\,0.1$ \cite{Isella2016}. For the starting point, we did not introduce any gap, and using other 
forms will not have a significant impact to the final result.
\item[b)] With an initial guess for $H_{\rm gas}$, the dust density distribution is given by Eq.~\ref{eqn:sgp} and \ref{eqn:lgp}.
          Radiative transfer modeling is performed to obtain the dust temperature. Then, the dust density structure is solved assuming that the disk 
          is in vertical hydrostatic equilibrium. We run the radiative transfer modeling with the new dust density distribution to get the new dust temperature. 
		  The iteration for the dust temperature and density goes back and forth, and convergence can be achieved after ${\sim}\,5$ iterations. 
		  For the initial choice of $H_{\rm gas}$, we assume $H_{\rm gas}\,{=}\,\sqrt{kT(R)R^3/GM_{\star}{\mu}m_{p}}$, where $G$ is the gravitational constant, 
		  $k$ is the Bolzmann's constant, $m_{p}$ is the mass of proton, $\mu\,{=}\,2.3$ is the mean molecular weight, and 
		  $T(R)=18.7(R/400\,\rm{AU})^{-0.14}$ is the midplane temperature given by Dullemond et al. \cite{Dullemond2020}. The black solid line in 
		  Figure~\ref{fig:s2hgas} shows the initial $H_{\rm gas}$. This step is time consuming because a smooth temperature structure is required to get the 
		  solution for the corresponding dust density. Thus, we use a total number of $3\,{\times}\,10^7$ photons in the simulation. 		  
\item[c)] From step b), the gas scale height ($H_{\rm gas}$) is derived self-consistently. Then, we simulate a model image at 1.25\,mm, which is convolved 
          with the ALMA beam that has a size and position angle of $0.048^{\prime\prime}\times0.038^{\prime\prime}$ and $82^{\circ}$, respectively.
\item[d)] We extracted the model surface brightness along the disk major axis to the northwest, identical to what we have done on the ALMA image.
\item[e)] A ratio as a function of radius $\zeta(R)$ is obtained by dividing the observed brightness profile by the model brightness profile. 
\item[f)] The surface densities used as the input for the model is scaled by the point-by-point ratios $\zeta(R)$. The process goes back to step b.
\end{itemize}

The iteration for $\Sigma_{\rm d}$ typically converges after about 25 loops, when the change in the model brightness profile is less than 5\% at all radii.

\begin{figure}[H]
 \centering
 \includegraphics[width=0.48\textwidth]{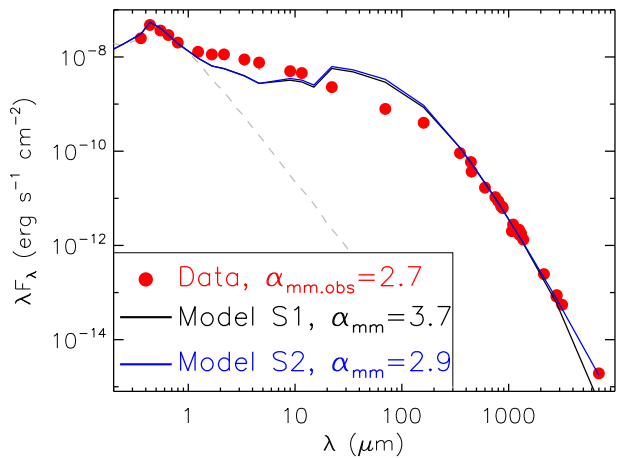}
 \caption{SED of the HD\,163296 disk. Red dots indicate photometric data that are taken from literature. The black and blue lines show two models with 
 the maximum grain size being $a_{\rm max}\,{=}\,1\,\rm{mm}$ and 1\,cm, respectively. The grey dashed line denotes the photospheric spectrum. The spectral 
 indices measured at wavelengths $\lambda\,{\ge}\,1\,\rm{mm}$ are given for both the models and observation.}
 \label{fig:bestsed}
\end{figure}

\begin{table*}[!t]
\centering	
\footnotesize
\begin{threeparttable}
\caption{Overview of parameter values for different models.}
\label{tab:paras}
\doublerulesep 0.1pt \tabcolsep 7pt %space between two columns. 
\linespread{1.2}\selectfont
\begin{tabular}{lcccccccl}
\toprule
 Parameter                &     Fixed/free    &  Model \texttt{S1}   &  Model \texttt{S2}   &   Model \texttt{I1}  &  Model \texttt{I2}  &  Model \texttt{I3}  &  Model \texttt{I4}  &  Note      \\
 \hline
 $T_{\rm eff}$\,[K]       &   Fixed     & \multicolumn{6}{c}{9250}            &   Effective temperature  \\
 $L_{\star}\,[L_{\odot}]$ &   Fixed     & \multicolumn{6}{c}{17}              &   Stellar luminosity  \\
 $D$\,[pc]                &   Fixed     & \multicolumn{6}{c}{101}             &   Distance  \\
 $i\,[^{\circ}]$          &   Fixed     & \multicolumn{6}{c}{46.7}            &  Disk inclination \\
 ${\rm PA\,[^{\circ}]}$   &   Fixed     & \multicolumn{6}{c}{133.33}          &  Position angle \\
 $R_{\rm in}$\,[AU]       &   Fixed     & \multicolumn{6}{c}{0.4}             &   Disk inner radius  \\
 $R_{\rm out}$\,[AU]      &   Fixed     & \multicolumn{6}{c}{169}             &   Disk outer radius  \\
 $f_{\rm LGP}$            &   Fixed     & \multicolumn{6}{c}{0.85}            &   Mass fraction of the LGP  \\
 $a_{\rm min.SGP}$\,[$\mu{\rm m}$]&   Fixed  &  \multicolumn{6}{c}{0.01}      &  Minimum grain size for the SGP \\
 $a_{\rm max.SGP}$\,[$\mu{\rm m}$]&   Fixed  &  \multicolumn{6}{c}{2}         &  Maximum grain size for the SGP \\
 $a_{\rm min.LGP}$\,[$\mu{\rm m}$]&   Fixed  &  \multicolumn{6}{c}{2}         &  Minimum grain size for the LGP \\
 $a_{\rm max.LGP}$\,[cm]  &   Free      &  0.1      &  1.0         & 1.0      &  1.0         & 1.0           &  1.0    &  Maximum grain size for the LGP \\
 $\Sigma_{\rm d}\,\rm{[g\,cm^{-2}]}$ &  Free  &  Figure~\ref{fig:s1s2surdens}  &  Figure~\ref{fig:s1s2surdens}  & Figure~\ref{fig:surdensout}  &  Figure~\ref{fig:surdensout}  & Figure~\ref{fig:surdensout}    &  Figure~\ref{fig:surdensout}   &  Dust surface density \\     
 $M_{\rm dust}\,[10^{-4}\,M_{\odot}]$ \tnote{(a)}    &   $-$    &  1.2         &  2.4     & 2.3        &  2.4           & 2.4             &  2.5      &  Total dust mass \\     
 $\Lambda$    &   Free    &  5.0     &  5.0         & 1.0       &  2.6           & 10.6             &  $-$      & $\Lambda$ for the entire disk, see Sect.~\ref{sec:conhratio} \\  
 $\Lambda{1}$    &   Free    &  $-$     &  $-$         & $-$        &  $-$           & $-$             &  $3.0_{-0.8}^{+0.3}$    &  $\Lambda$ for ${R\,{<}\,59\,\rm{AU}}$, see Sect.~\ref{sec:varhratio} \\  
 $\Lambda{2}$    &   Free    &  $-$     &  $-$         & $-$        &  $-$           & $-$             &  $1.2_{-0.1}^{+0.1}$    &  $\Lambda$ for ${59\,{\le}\,R\,{<}\,78\,\rm{AU}}$, see Sect.~\ref{sec:varhratio} \\  
 $\Lambda{3}$    &   Free    &  $-$     &  $-$         & $-$        &  $-$           & $-$             &  $1.9_{-0.1}^{+15.9}$   &  $\Lambda$ for ${78\,{\le}\,R\,{<}\,94\,\rm{AU}}$, see Sect.~\ref{sec:varhratio} \\  
 $\Lambda{4}$    &   Free    &  $-$     &  $-$         & $-$        &  $-$           & $-$             &  $16.3_{-9.8}^{+3.7}$   &  $\Lambda$ for ${R\,{\ge}\,94\,\rm{AU}}$, see Sect.~\ref{sec:varhratio} \\ 
 \hline
 $\chi_{\rm tot}^2$    &   $-$    &  $-$   &  $-$     & 975        &  460           & 478             &  242   &  Chi-square of the model, see Eq.~\ref{eqn:chitot} \\  
\bottomrule
\end{tabular}
\begin{tablenotes}
\item[(a)] The total dust mass $M_{\rm dust}$ is obtained by integrating the surface density $\Sigma_{\rm d}$ that 
is constructed in the fitting procedure. Hence, $M_{\rm dust}$ is not a direct fitting parameter.
\end{tablenotes}
\end{threeparttable}
\end{table*}

\subsection{Setting $a_{\rm{max}}$ for the LGP based on SED modeling}
\label{sec:sedmodel}

Our model has three free parameters/quantities: the dust surface density ($\Sigma_{\rm d}$), the ratio of gas-to-dust scale height ($\Lambda$) and 
maximum grain size ($a_{\rm max}$) for the LGP. Note that the total dust mass ($M_{\rm dust}$) is not a free parameter, because integrating $\Sigma_{\rm d}$
within the disk naturally gives the result. 

A population of large dust grains will shallow the spectral index at millimeter wavelengths \cite{Ricci2010,Testi2014}. We collected photometric data from 
various catalogs and individual studies \cite{Mannings1994,Oudmaijer2001,cutri2003,Isella2007,ishihara2010,Sandell2011,cutri2013,Pascual2016,Tripathi2017,Andrews2018,Guidi2022}. 
The observed spectral energy distribution (SED) is shown as red dots in Figure~\ref{fig:bestsed}. The spectral index measured at wavelengths $\lambda\,{\ge}\,1\,\rm{mm}$ 
is $\alpha_{\rm mm.obs}\,{=}\,2.7\,{\pm}\,0.06$. Assuming that the emission is optically thin and in the Rayleigh-Jeans tail, this transfers into a millimeter 
slope of the dust absorption coefficient $\beta\,{=}\,\alpha_{\rm mm.obs}{-}\,2\,{=}\,0.7$. The $\beta$ value for the interstellar medium dust 
is ${\sim}\,1.7$ \cite{Li2001}. A lower $\beta$ in the HD\,163296 disk suggests that dust grains have grown up to millimeter and even centimeter sizes.  

To quantify the extent of grain growth in the HD\,163296 disk, we build a grid of SED models in which the ratio of gas-to-dust scale height is fixed to 
$\Lambda\,{=}\,5$, a typical value used in literature works \cite{andrews2011,Liu2022}. In Sect.~\ref{sec:fitalma}, we will conduct an extensive parameter
study on $\Lambda$ through a dedicated fitting to the ALMA image. However, this parameter is not expected to have a significant impact to the constraint on 
$a_{\rm max}$ as long as the optical depth is not large. We sample 16 different $a_{\rm max}(s)$ that are logarithmically distributed from $10\,\mu{\rm m}$ to 1\,cm. 
The procedure of iteration for $\Sigma_{\rm d}$, as laid out in Sect.~\ref{sec:surdens}, is performed separately for each of the 16 models. As a result, 16 model 
SEDs are simulated. The model with $a_{\rm max}\,{=}\,1\,\rm{cm}$ (Model \texttt{S2}) best matches with the observation, see Figure~\ref{fig:bestsed}. Its converged 
surface density is shown in Figure~\ref{fig:s1s2surdens}, and Table~\ref{tab:paras} gives an overview of the model parameters. For the subsequent fitting 
to the ALMA data, we fixed $a_{\rm max}\,{=}\,1\,\rm{cm}$ for the LGP, leaving $\Sigma_{\rm d}$ and $\Lambda$ as the only two free parameters. 

The discrepancies in the mid- and far-infrared fluxes between model and observation is due to the presence of a puffed-up inner rim. This type of rim is a natural 
outcome when solving the disk structure in vertical hydrostatic equilibrium, particularly for Herbig disks \cite{Dullemond2007}. The blue solid line in 
Figure~\ref{fig:s2hgas} shows the gas scale height of Model \texttt{S2}. The overall geometry of the disk is flared. Disk regions just behind the inner rim 
cannot be exposed to the stellar light, leading to a reduced mid-infrared excess. At a certain radial distance, the disk will show up from the shadow 
casted by the inner rim. The surface layer of these outer regions directly absorbs stellar photons, and hence produces more far-infrared emission than 
the observed level. One can fully parameterize the scale height with analytic forms, e.g., a power law, and fit the infrared SED to constrain the 
geometry \cite{Harvey2012}. However, there are some degeneracies between the geometric parameters in SED models. Moreover, modeling the SED is not 
able to constrain the scale height of millimeter dust grains that is the key of this work. Therefore, we do not attempt to conduct further fine tuning 
on the SED fitting, and make our assumptions (i.e., number of free parameters) as few as possible.

\section{Fitting the DSHARP ALMA image}
\label{sec:fitalma}

In this section, we will fit the surface brightnesses along the major and minor axes of the disk, and on the B67 and B100 rings to 
constrain $\Lambda$. Our strategy starts from a simple assumption of a constant $\Lambda$ in the radial direction, to a more complex 
scenario in which $\Lambda$ varies with $R$. 

The contrasts of gaps, as presented in Table~\ref{tab:gapcont}, are sensitive to the degree of dust settling. Therefore, to quantify
the quality of fit, we first check whether or not the gap contrasts of the model are consistent with the observation. Then, we 
calculate the $\chi^2$ along the major axis ($\chi_{\rm major}^2$) and minor axis ($\chi_{\rm minor}^2$), and on the B67 ($\chi_{\rm B67}^2$) 
and B100 ring ($\chi_{\rm B100}^2$). To exclude the effect of the crescent-like substructure along ${\rm PA}\,{\sim}\,99^{\circ}$, 
data points between $\phi\,{=}\,{-}45^{\circ}$ and $45^{\circ}$ are not taken into account when calculating $\chi_{\rm B67}^2$ and $\chi_{\rm B100}^2$. 
The goodness of fit is evaluated according to 

\begin{equation}
\chi_{\rm tot}^2\,{=}\,g_{1}\,\chi_{\rm major}^2+g_{2}\,\chi_{\rm minor}^2+g_{3}\,\chi_{\rm B67}^2+g_{4}\,\chi_{\rm B100}^2.
\label{eqn:chitot}
\end{equation}
Four factors, i.e., $g_{1}$, $g_{2}$, $g_{3}$ and $g_{4}$, are introduced to balance the weightings. First, we calculate the factors as 
\begin{equation}
 g_{i} = \frac{\sum_{i=1}^{4}\left(N_{i}\right)}{N_{i}},
\end{equation}
where $N_{i}$ is the number of data points taken into account in the calculation of $\chi^2(s)$ for the major and minor axes, and the B67 and B100 
rings, respectively. Then, a normalization is performed to ensure that the sum of $g_{i}$ equals to unity.

\subsection{Constant $\Lambda$ in the radial direction} 
\label{sec:conhratio}

We first take the simplest assumption in which the ratio of gas-to-dust scale height does not change with radius ($R$). 
We sample 20 values for $\Lambda$, which are logarithmically distributed within 1 and 20. The case of $\Lambda\,{=}\,1$ 
means that millimeter dust grains are well coupled with the gas. Strongly settled models feature large values of $\Lambda$. The iteration 
process for $\Sigma_{d}$ is performed from scratch for each of these 20 models, ensuring that all the models are fully independent 
and self-consistent. 

\begin{figure}[H]
 \centering
 \includegraphics[width=0.48\textwidth]{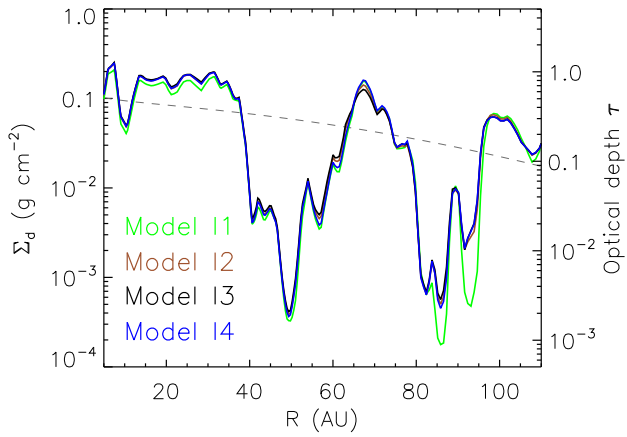}
 \caption{Dust surface densities (on the left Y axis) reconstructed from the iterative fitting process, and optical depth at a wavelength of 
 1.25\,mm (on the right Y axis) for model I1, I2, I3 and I4. The dashed line shows the starting surface density used in the fitting 
 loop: $\Sigma(R)\,{=}\,\Sigma_{0}\,(R/R_{\rm c})^{-\gamma}\,{\rm exp}[-(R/R_{\rm c})^{2-\gamma}]$ with $R_{\rm{c}}\,{=}\,90\,\rm{AU}$ 
 and $\gamma\,{=}\,0.1$, see Sect.~\ref{sec:surdens}.}
\label{fig:surdensout}
\end{figure}

None of the 20 models can reproduce all of the gap constrasts within the uncertainties simultaneously. Panels (c)-(f) of Figure~\ref{fig:imgres} 
shows a comparison of the brightnesses between observation and three representative models with $\Lambda\,{=}\,1.0$ (model \texttt{I1}), 
2.6 (model \texttt{I2}) and 10.6 (model \texttt{I3}), respectively. Model \texttt{I2} has the lowest $\chi_{\rm tot}^2\,{=}\,460$ among the 20 samples. 
Figure~\ref{fig:surdensout} shows the reconstructed surface densities, whereas the gap contrasts extracted from the models 
are given in Table~\ref{tab:gapcont}. 

Along the disk major axis, the three models reproduce the data at a similar quality, see panel (c) of Figure~\ref{fig:imgres},
and the model gap contrasts in Table~\ref{tab:gapcont}. The ALMA beam can dilute the ring emission, and contributes to the 
adjacent gap emission. In vertically thicker (smaller $\Lambda$) disks, dust grains are located at a higher height above the 
midplane where the temperature is high. In this case, the ring emission is stronger, and its contribution to the gap emission 
is higher, which can shallow the gap contrast since the intrinsic emission from the gap is low. In addition to the millimeter dust 
scale height, the depth of surface density drops is another quantity influencing the gap contrast. A comparison between 
model \texttt{I1} and model \texttt{I3} indicates that deeper surface density drops in more turbulent disks can produce similar 
gap contrasts measured on the disk major axis to those generated by shallower surface density drops in more quiescent 
disks. This means that fitting the data on the major axis alone cannot break the degeneracy. 
 
Along the disk minor axis, the change to the gap contrast as a function of $\Lambda$ is observed due to the effect of projection.  
Panel (d) of Figure~\ref{fig:imgres} shows that models with a higher degree of dust settling produce more separate rings and deeper gaps, 
and vice versa. This fact is consistent with the findings reported by Pinte et al. \cite{Pinte2016}. Neither the D48 nor the D86 gap 
can be explained by model \texttt{I1}. Though both model \texttt{I2} and \texttt{I3} are consistent with the data of the D48 gap, only the 
former reproduces the D86 gap within the uncertainty, see Table~\ref{tab:gapcont}.
  
The gas-to-dust scale height ratio $\Lambda$ has a strong impact on the brightness variation on the B67 and B100 rings. The well-mixed disk 
(model \texttt{I1}) shows two pronounced dips at $\phi\,{=}\,90^{\circ}$ and $\phi\,{=}\,270^{\circ}$, due to the difference in the optical depth ($\tau$) along 
the line of sight between $\phi\,{=}\,0^{\circ}$ (or $180^{\circ}$, major axis) and $\phi\,{=}\,90^{\circ}$ (or $270^{\circ}$, minor axis) \cite{Doi2021}. 
Such a difference in $\tau$ decreases with increasing $\Lambda$. Consequently, the contrasts of ``gaps'' on the rings are reduced in more settled 
disks, see for instance model \texttt{I3}. Panels (e) and (f) of Figure~\ref{fig:imgres} suggest that the degree of dust settling is different 
between B67 and B100. While B67 is close to a well-mixed situation, B100 favors a scenario in which large dust grains are well concentrated 
in the midplane. 

\subsection{Varying $\Lambda$ in the radial direction} 
\label{sec:varhratio}
Though the experiment under the assumption for a constant $\Lambda$ does not return a satisfactory solution, it provides clues to improve the model. 
The fitting results imply that the degree of dust settling changes with $R$. Therefore, we parameterize 
the ratio of gas-to-dust scale height with a piecewise function
\begin{equation}
 \Lambda = \left\{
 \begin{array}{rcl}
	 \Lambda{1} & : & {R\,{<}\,59\,\rm{AU}} \\
	 \Lambda{2} & : & {59\,\rm{AU}\,{\le}\,R\,{<}\,78\,\rm{AU}} \\
	 \Lambda{3} & : & {78\,\rm{AU}\,{\le}\,R\,{<}\,94\,\rm{AU}} \\
	 \Lambda{4} & : & {R\,{\ge}\,94\,\rm{AU}}. \\
 \end{array} \right.	
\end{equation}
The boundaries of the four radial bins are chosen according to the locations and widths of the gaps and rings, see Sect.~\ref{sec:obs}. 
We did not explore these borders in the fitting process. Using a piecewise form may have some artifacts in the boundaries. Nevertheless, how the 
gas-to-dust scale height ratio smoothly varies from one radial bin to another is difficult to be investigated, because it requires observational data 
at extremely high spatial resolutions that fully resolve the transition region between two adjacent bins. In the new model configuration, the 
ratios $\Lambda{2}$ and $\Lambda{4}$ are expected to play the dominated role in controlling the gap contrasts of the B67 and B100 rings, respectively. 
The gap contrasts of D48 and D86 are mainly influenced by a combination of $\Lambda{1}$ and $\Lambda{2}$, and a combination of $\Lambda{3}$ 
and $\Lambda{4}$, respectively. This is because the definition of contrasts of gaps on the major/minor axis is related to the brightnesses both in 
the gap and in its exterior ring, see Sect.~\ref{sec:obs}.

\begin{figure*}[t]
 \centering
 \includegraphics[width=\textwidth]{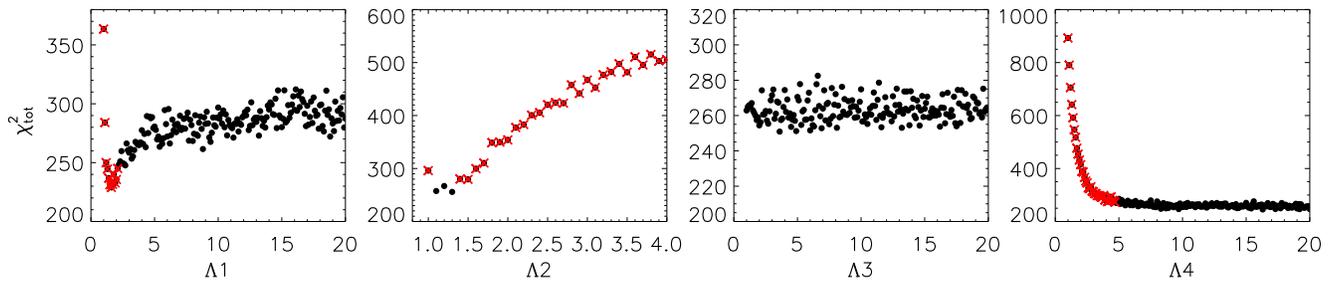}
 \caption{The $\chi_{\rm tot}^2$ distribution as a function of the gas-to-dust scale height ratio $\Lambda{1}$, $\Lambda{2}$, $\Lambda{3}$ and $\Lambda{4}$.
          The dots overlaid with a red cross refer to models that cannot reproduce all of the observed gap contrasts tabulated Table~\ref{tab:gapcont}. 
		  Note that the $\chi_{\rm tot}^2\,{-}\,\Lambda{3}$ profile is flat, and most of the considered values for $\Lambda{3}$ are able to generate all 
		  of the observed gap contrasts. Therefore, $\Lambda{3}$ is basically not constrained.}
\label{fig:chi2tot}
\end{figure*}

The parameter space becomes $\left\{\Lambda{1},\,\Lambda{2},\,\Lambda{3},\,\Lambda{4},\,\Sigma_{d}\right\}$. To maintain self-consistency 
and independency, the time-consuming process for iterating $\Sigma_{d}$ has to be conducted for each of the sampled sets $\left\{\Lambda{1},\,\Lambda{2},\,\Lambda{3},\,\Lambda{4}\right\}$. 
Therefore, it is impractical to perform the parameter study using the Markov Chain Monte Carlo approach. Instead, the grid search method is 
invoked to finish the task. We first search for the optimum combination of $\Lambda{1}$ and $\Lambda{2}$, and then for that of $\Lambda{3}$ 
and $\Lambda{4}$. We sample 20 values for $\Lambda{1}$, which are logarithmically spaced from 1 and 20. Before the parameter study, we run 
many simulation tests, and find that models with $\Lambda{2}$ only slightly deviating from ${\sim}\,1.2$ are not able to generate gap 
contrasts of B67 comparable to the observation. Hence, for the sake of reducing the computational time and meanwhile being conservative, we consider 
10 points for $\Lambda{2}$ from 1 to 4 in the logarithmic manner. At this stage, $\Lambda{3}$ and $\Lambda{4}$ are fixed to 2.6, i.e., the 
value of model \texttt{I2}. We run the iteration procedure for $\Sigma_{d}$ from scratch for each of the 200 different combinations of $\Lambda{1}$ 
and $\Lambda{2}$, and obtain 200 models. Then, we fix $\Lambda{1}$ and $\Lambda{2}$ to the values of the model with the lowest 
$\chi_{\rm tot}^2$. The exploration for $\Lambda{3}$ and $\Lambda{4}$ is similar. However, both parameters have the same grid points to those 
for $\Lambda{1}$, and therefore they form 400 different combinations. 

The final best-fit model (model \texttt{I4}) features $\Lambda{1}\,{=}\,3.0$, $\Lambda{2}\,{=}\,1.2$, $\Lambda{3}\,{=}\,1.9$, $\Lambda{4}\,{=}\,16.3$, 
and $\chi_{\rm tot}^2\,{=}\,245$. Its dust surface density and millimeter optical depth are shown with the blue line in Figure~\ref{fig:surdensout}. 
The model image and brightness profiles are compared with the observation in Figure~\ref{fig:imgres}. The gap contrasts and model parameters are summarized 
in Table~\ref{tab:gapcont} and \ref{tab:paras}, respectively. The best-fit model is able to explain all of the gap contrasts. 
We separately vary the gas-to-dust scale height ratios in each radial bin from their best-fit values with a step width of 0.1, and investigate how 
well the parameters are constrained. The variations of $\chi_{\rm tot}^2$ are shown in Figure~\ref{fig:chi2tot}. The dots overlaid with a red cross 
refer to models that cannot reproduce all of the observed gap contrasts within their errors. Therefore, we exclude them 
in the estimation of parameter uncertainties that are deduced from the models with $\chi_{\rm tot}^2$ less than 1.05 times the minimum $\chi_{\rm tot}^2$. 
For instance, all the models with $\Lambda1\,{<}\,{\sim}2.2$ produce lower contrasts (i.e., ${<}\,0.92$) for the D48 gap measured on the disk minor axis 
than the observed value ($0.94\,{\pm}\,0.02$). Therefore, they are considered to be invalid although some of them have better $\chi^2_{\rm tot}$ than that of the 
best-fit model. The profiles of $\chi_{\rm tot}^2\,{-}\,\Lambda{1}$, $\chi_{\rm tot}^2\,{-}\,\Lambda{2}$ and $\chi_{\rm tot}^2\,{-}\,\Lambda{4}$ show
a clear signature of getting the optimum solution, indicating that the gas-to-dust scale height ratios in the D48, B67 and B100 regions are well constrained. 
Their validity ranges are estimated to be [2.2, 3.3], [1.1, 1.3], and ${\ge}\,6.5$, respectively. The distribution of $\chi_{\rm tot}^2$ as a function of $\Lambda{3}$ 
is quite flat, and all the $\Lambda{3}$ values in the considered range can reproduce the data well. Hence, $\Lambda{3}$ is basically unconstrained.

\section{Discussion}
\label{sec:discussion}

Using self-consistent radiative transfer models, we have placed constraints on the degree of dust settling by fitting the 
gap contrasts of the D48, B67, D86 and B100 features. Our results suggest a radially varying ratio of gas-to-dust scale 
height ratio in the HD\,163296 disk. In this section, we compare our result with literature studies, and link the 
derived gas-to-dust scale height ratio to the turbulence strength in the HD\,163296 disk.

\subsection{Comparison of $\Lambda$ between different works}

\begin{figure}[H]
 \centering
 \includegraphics[width=0.48\textwidth]{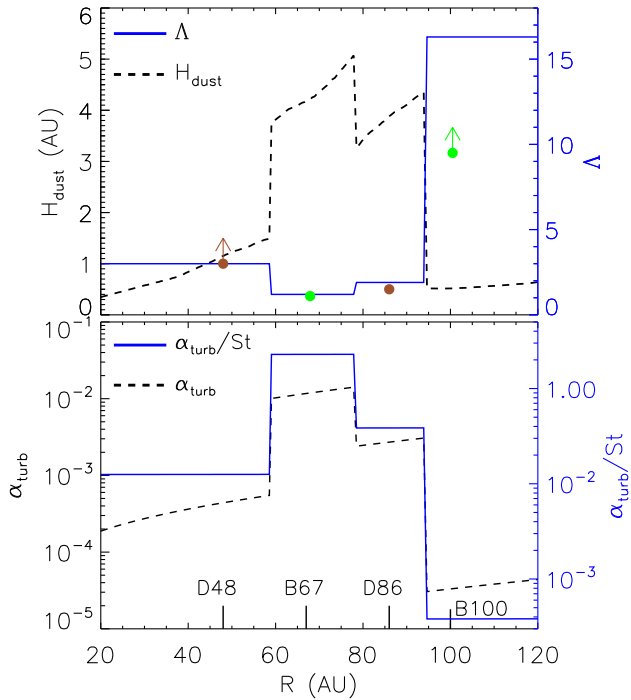}
 \caption{{\it Upper panel:} comparison of $\Lambda$ (on the right Y axis) between model \texttt{I4} (blue solid line) and literature studies. 
 The $\Lambda$ values by Ohashi et al. \cite{Ohashi2019} and Doi \& Kataoka \cite{Doi2021} are indicated with brown dots and green dots, respectively. 
 The black dashed curve shows the millimeter dust scale height (on the left Y axis) of model \texttt{I4}. {\it Bottom panel:} the  
 $\alpha_{\rm turb}/{\rm St}$ ratio (blue solid line, on the right Y axis) and $\alpha_{\rm turb}$ (black dashed line, on the left Y axis) of model \texttt{I4}.}
\label{fig:hdustcompare}
\end{figure}

Ohashi et al. \cite{Ohashi2019} found that the dust scale height is the key parameter for reproducing the azimuthal variation of the
polarization pattern in the gaps. By analyzing the ALMA data of the 0.87\,mm dust polarization from the HD\,163296 disk, they constrained 
the dust scale height to be less than one-third the gas scale height for the D48 gap, and to be two-thirds the gas scale height for the 
D86 gap.  Recently, Doi \& Kataoka \cite{Doi2021} showed that the azimuthal variation in the continuum along rings are sentitive 
to the degree of dust settling. Assuming that the disk is vertically isothermal with a fixed power-law temperature, they 
fit the DSHARP continuum data of the B67 and B100 rings, and inferred the ratio of gas-to-dust scale height to be 1.1 and ${>}\,9.5$ for 
the B67 and B100 ring, respectively. Figure~\ref{fig:hdustcompare} (upper panel) shows a comparison of $\Lambda$ between different works. 
The blue solid line refers to our best fit, whereas brown dots and green dots mark the results by Ohashi et al. \cite{Ohashi2019} 
and Doi \& Kataoka \cite{Doi2021}, respectively. As can be seen, our results are overall consistent with these literature values. However, 
as one step further, our analysis provides constraints on $\Lambda$ both for the ring and gap regions in the framework of self-consistent 
radiative transfer simulation.

The black dashed line in the upper panel of Figure~\ref{fig:hdustcompare} shows the dust scale height. In the inner ($R\,{<}\,60\,\rm{AU}$) 
or outermost ($R\,{>}\,94\,\rm{AU}$) regions, the millimeter dust disk is quite thin, with scale heights less than ${\sim}\,2\,\rm{AU}$. 
Disk regions in the vicinity of B67 have millimeter dust scale height of ${\sim}\,4\,\rm{AU}$. Disks, when viewed at high 
inclinations, have a specific advantage that the vertical extent of the emission layers can be directly constrained by spatially 
resolved images. Villenave et al. \cite{Villenave2020} presented ALMA continuum observations of 12 edge-on disks, at an angular 
resolution of ${\sim}\,0.1^{\prime\prime}$. A comparison between a set of radiative transfer models and the data indicates that at 
least three disks in their sample are consistent with a millimeter dust scale height of a few AU. Our inferred dust scale height 
for the HD\,163296 disk, tilted to $46.7^{\circ}$, is comparable with those of the observed edge-on disks.

\subsection{Comparison of $\alpha_{\rm turb}/\rm{St}$ and $\alpha_{\rm turb}$ between different works}

Assuming an equilibrium between dust settling and vertical stirring by turbulent motions, the dust scale height and gas scale height follow  
the relation \cite{Youdin2007,Birnstiel2010} 
\begin{equation}
 H_{\rm dust} = H_{\rm gas}\left(1+\frac{\rm St}{\alpha_{\rm turb}} \frac{\rm 1+2\,St}{\rm 1+St}\right)^{-1/2},
\label{eqn:dusth}
\end{equation}
where the Stokes number St is given by 
\begin{equation}
 {\rm St}\,{=}\,\frac{\rho_{\rm grain}\bar{a}}{\Sigma_{\rm g}(R)}\frac{\pi}{2}. 
\end{equation}
The gas surface density $\Sigma_{\rm g}(R)\,{=}\,\Sigma_{0}\,(R/R_{\rm c})^{-\gamma}\,{\rm exp}[-(R/R_{\rm c})^{2-\gamma}]$ 
with $\Sigma_{0}\,{=}\,8.8\,\rm{g\,cm^{-2}}$, $R_{\rm{c}}\,{=}\,165\,\rm{AU}$ and $\gamma\,{=}\,0.8$, are constrained by high resolution multiple CO line 
observations \cite{Zhang2021}. Considering a grain size distribution like the one prescribed for the LGP, $\bar{a}$ stands for the representative grain 
size of dust that dominates the continuum emission at 1.25\,mm. We check how the mass absorption coefficent $\kappa_{\rm abs}$ at 1.25\,mm changes 
with $a$, and find that it peaks at $a\,{\sim}\,0.2\,\rm{mm}$. This value is close to the number given by $\lambda/2\pi$. Therefore, in our calculation
of St, we took $\bar{a}\,{=}\,0.2\,\rm{mm}$. 

\begin{table}[H]
\caption{$\alpha_{\rm turb}/{\rm St}$ for the B67 and B100 ring from different studies.}
\centering
\linespread{1.2}\selectfont
\begin{tabular}{lcc}
\hline
Reference                             & B67 ring              &     B100 ring   \\
\hline
Dullemond et al. \cite{Dullemond2018} & 0.33                  &     $0.13\,{\sim}\,0.77$ \\
Rosotti et al. \cite{Rosotti2020}     & 0.23                  &     0.04 \\
Doi \& Kataoka \cite{Doi2021}         & ${>}\,2.4$            &     ${<}\,0.011$ \\
This work                             & $2.3_{-0.9}^{+2.5}$   &    $0.0038_{-0.0013}^{+0.02}$  \\   
\hline
\end{tabular}
\linespread{1.0}\selectfont
\label{tab:alpha}
\end{table}

The St value varies from ${\sim}\,10^{-5}$ in the inner disk to ${\sim}\,10^{-2}$ in the outer regions. Because St is much less than unity, Eq.~\ref{eqn:dusth} 
can be simplified as $H_{\rm dust}\,{=}\,H_{\rm gas}\left(1+\frac{\rm St}{\alpha_{\rm turb}}\right)^{-1/2}$. Therefore, the constrained $\Lambda$ 
directly translates into a ratio of $\alpha_{\rm turb}/\rm{St}$, which is shown with the blue solid line in the bottom panel of Figure~\ref{fig:hdustcompare}.
Based on different methodologies, other groups have derived the $\alpha_{\rm turb}/\rm{St}$ values for the B67 and B100 rings. For instance, Rosotti et al. \cite{Rosotti2020} 
determined $\alpha_{\rm turb}/\rm{St}$ by measuring the deviation from Keplerian rotation of the gas in the proximity of the continuum peaks. Under an 
assumption that dust rings are caused by dust trapping in radial pressure bumps, Dullemond et al. \cite{Dullemond2018} constrained $\alpha_{\rm turb}/{\rm St}$ 
by analyzing the widths of the dust rings. In Doi \& Kataoka \cite{Doi2021}, the $\alpha_{\rm turb}/\rm{St}$ value was inferred by investigating the azimuthal 
intensity variation along dust rings. Table~\ref{tab:alpha} summarizes the reported values together with our best-fit result. As can be seen, our result
is well consistent with the values derived by Doi \& Kataoka \cite{Doi2021}. This is not surprising because the idea of constraning $\alpha_{\rm turb}/\rm{St}$
is the same. But, our methodology is more realistic, and data points not only on the rings but also along the major/minor axes are simultaneously taken into account 
in the analysis. We note that the best-fit $\alpha_{\rm turb}/{\rm St}$ for B67 is about one order of magnitude larger than those obtained in Dullemond et al. and 
Rosotti et al. There are several possibilities to explain such a difference. First, our methodology is sensitive to the strength of turbulent motions in the vertical 
direction, while the constraints by Dullemond et al. are more related to the radial diffusion of dust grains. Second, the B67 ring has a neighboring crescent, implying 
that the ring itself may not be perfectly axisymmetric, thus undermining the assumption of our modeling procedure. Third, if the gaps are indeed opened by 
planets \cite{Pinte2018,Teague2018b,Teague2021}, the B67 ring can be substantially stirred due to meridional gas flows. Numerical simulations have shown that massive 
planets can stir sub-millimeter-sized dust grains up to ${\sim}\,70\%$ of the gas scale height at the gap edges \cite{Bi2021,Binkert2021}. For the B100 ring, we obtain 
a lower $\alpha_{\rm turb}/{\rm St}$ than that inferred by Dullemond et al. A lower turbulence in the vertical direction than in the radial direction can be explained 
under several physical scenarios, such as in dust feedback to turbulence \cite{Xu2022}, disk self-gravity \cite{Baehr2021}, and radial (pseudo-)diffusion \cite{Hu2021}.

The black dashed line in the bottom panel of Figure~\ref{fig:hdustcompare} shows the derived turbulence strength. Except for 
the B67 ring, the disk has a turbulence level of $\alpha_{\rm turb}\,{<}\,3\times10^{-3}$. Theoretical works
have shown that pure hydrodynamic mechanisms or the magnetorotational instability suppressed by nonideal magnetohydrodynamic effects 
can generate similar turbulence levels in protoplanetary disks \cite{Bai2011,Bai2015,Flock2017,Cui2020,Cui2021}.
In the B67 ring, the turbulence is strong with $\alpha_{\rm turb}\,{\sim}\,1.2\,{\times}\,10^{-2}$.

Several studies have tried to measure turbulence in the HD\,163296 disk through detailed analysis of gas line observations. Boneberg et al. \cite{Boneberg2016} 
found that models with $\alpha_{\rm turb}\,{=}\,(0.1\,{-}\,6.3)\,{\times}\,10^{-3}$ match well with the ${\rm C^{18}O}\,J\,{=}\,2{-}1$ line 
profile within 90\,AU of the disk. Based on CO isotopes and DCO$^{+}$ line observations, Flaherty et al. \cite{Flaherty2015,Flaherty2017} 
derived the gas turbulence velocity in the disk, which is less than a few percent of the sound speed, corresponding 
to $\alpha_{\rm turb}\,{<}\,{\sim}\,3\,{\times}\,10^{-3}$. Our inferred value for $\alpha_{\rm turb}$, except for the B67 ring, is 
consistent with the results set by gas observations. The value of $\alpha_{\rm turb}$ for B67 from our modeling is larger 
than the upper limit in either Boneberg et al. or Flaherty et al. The discrepancy may be explained by two reasons. First, 
as demonstrated by our analysis, $\Lambda$, and therefore $\alpha_{\rm turb}$, may vary in the radial direction. The spatial resolution 
of gas line observations in Boneberg et al. and Flaherty et al. is ${\sim}\,0.5^{\prime\prime}$ that is 10 times worse than that of 
the DSHARP data. Consequently, their constraints on $\alpha_{\rm turb}$ represent a mean level of turbulence over a much broader 
range of radius than ours. Due to the beam smearing, low turbulence outside B67 results in a small $\alpha_{\rm turb}$ probed 
by the gas lines. Second, the turbulence strength we measure describes the role of dust stirring in the vertical direction. This may be different from 
the turbulence of gas motions. Recent numerical simulations of dust evolution start to use different $\alpha_{\rm turb}(s)$ for gas 
evolution, radial diffusion and vertical stirring \cite{Pinilla2021}.
 
Isella et al. \cite{Isella2016} presented Band 6 ALMA observations of HD\,163296 with a lower angular resolution than the DSHARP data, revealing 
three dust gaps at 60, 100, and 160\,AU in the continuum as well as CO depletion in the middle and outer dust gaps. 
Liu et al. \cite{Liu2018} investigated these gaps by performing 2D global hydrodynamic simulations of planet-disk interaction, and found that 
three half-Jovian-mass planets in a disk with effective viscosity being a function of radius can explain most of the observational features. 
Within $R\,{=}\,100\,\rm{AU}$, their model has a turbulence level of $\alpha_{\rm turb}\,{<}\,3\,{\times}\,10^{-4}$ that is weaker than ours.
Such an inconsistency can be explained by the difference in the quality of data used in the analysis. As shown in the left column of Figure 3 in 
Liu et al. \cite{Liu2018}, the best-fit $\alpha_{\rm turb}$ is sensitive to how well the dust surface densities in the gap region are constrained. 
In the ALMA observation used by Liu et al. \cite{Liu2018}, the beam size is ${\sim}\,0.2^{\prime\prime}$ and the widths of the inner 
two gaps are narrower than ${\sim}\,0.27^{\prime\prime}$, indicating that the gaps are not fully resolved. However, our constraints are placed 
using the DSHARP data with four times better spatial resolution and sensitivity.

\subsection{The effect of model assumptions on the results}

%In our models, the dust scale height is prescribed based on the gas scale height according to Eq. \ref{eqn:dusth}. 
%We did not solve the vertical structure of the disk under the assumption of hydrostatic equilibrium, which means that 
%the gas scale height is fully parameterized (see Eq. \ref{eqn:gash}). This approach has been widely adopted in multi-wavelength 
%modeling of protoplanetary disks \cite{andrews2011,madlener2012,Liu2019}, and it has the advantage that one 
%can directly evaluate the effect of disk geometrical parameters on the result. Since the dust scale height of millimeter grains 
%is the driving factor for shaping the appearance of millimeter gaps, the gap contrast (therefore the inferred $H_{\rm 1mm.100au}$ 
%or $\alpha_{\rm turb}$) is expected to vary with $H_{100}$ or $\beta$. 

The direct constraint from our radiative transfer analysis is on the gas-to-dust scale height ratio $\Lambda$. The scenario of dust settling that 
links $\Lambda$ and $\alpha_{\rm turb}$ is given by Eq.~\ref{eqn:dusth}, and the relation is based on numerical simulations performed by Dubrulle et 
al. \cite{Dubrulle1995} and Youdin \& Lithwick \cite{Youdin2007}. Models with more realistic physics on dust growth, sedimentation and radial 
mixing may alter the connection between dust and gas scale heights, therefore change the result.

To calculate the Stokes number characterizing the coupling between gas and dust, one needs to know the gas surface density. In our calculation, 
we take the result from Zhang et al. \cite{Zhang2021} who modeled the high resolution ALMA data of CO and its isotopologue lines. How well
the CO molecular lines probe the underlying total gas surface density remains uncertain. Such a fact will not affect our constaints
on $\Lambda$ from the continuum radiative transfer modeling, but it will cause uncertainties when inferring $\alpha_{\rm turb}$ 
from $\Lambda$, see Eq.~\ref{eqn:dusth}. 

\section{Summary}
\label{sec:summary}
Constraining the strength of turbulence plays a key role in building up our knowledge on disk evolution and planet formation. 
It is also crucial for running numerical models to interpret high-resolution ALMA observations. %As an example, turbulence as an input for
%hydrodynamical simulation of planet-disk interaction highly affects the resulting depths and number of gaps that have been
%frequently observed in disks nowadays with ALMA.
In this work, we took the HD\,163296 disk as an example, and investigated in detail the millimeter gap contrast as a probe for turbulence 
level. With self-consistent radiative transfer modeling, we fit the gap contrasts measured for the D48, B67, D86 and B100 substructures that 
are spatially resolved by the DSHARP observation. We constrained the gas-to-dust scale height ratio $\Lambda$ to be $3.0_{-0.8}^{+0.3}$, $1.2_{-0.1}^{+0.1}$
and ${\ge}\,6.5$ for the D48, B67 and B100 regions. Our results show that the degree of dust settling varies with radius in the HD\,163296 disk. 
The $\Lambda$ value for the D86 region is unconstrained due to the degeneracy between $\Lambda$ and the depth of surface density drops.

Based on the constrained gas-to-dust scale height ratio $\Lambda$, we estimate $\alpha_{\rm turb}/\rm{St}$ to be $2.3_{-0.9}^{+2.5}$ 
and $0.0038_{-0.0013}^{+0.02}$ for the B67 and B100 rings, respectively. These values are well consistent with those reported by 
Doi \& Kataoka \cite{Doi2021}, but differ from the numbers inferred by Dullemond et al. \cite{Dullemond2018} and Rosotti et al. \cite{Rosotti2020}. 
The discrepancy may be due of the fact that our modeling is sentitive to the turbulence for vertical stirring of dust grains, while 
literature studies more likely reflect the turbulence for the radial diffusion of dust grains or the turbulent motion of gas species.  

We calculate the turbulence level to be $\alpha_{\rm turb}\,{<}\,3\times10^{-3}$ for the D48 and B100 regions, which agree well with the 
upper limit set by Boneberg et al. \cite{Boneberg2016} and Flarherty et al. \cite{Flaherty2017} from analyzing the width of gas
lines. According to our analysis, the B67 ring has a strong turbulence strength of $\alpha_{\rm turb}\,{\sim}1.2\,{\times}\,10^{-2}$.
Future multi-wavelength continuum observations with comparable spatial resolution to the DSHARP data are required to better 
constrain the degree of dust settling, and therefore the scale height of dust grains with different sizes. Higher resolution observations 
of multiple gas lines are pivotal to directly measure the turbulent motions, and confirm whether the strong turbulence in the local 
region of B67 inferred from our analysis is also seen with gas tracers.

%%%%%%%%%%%%%%%%%%%%%%%%%%%%%%%%%%%%%%%%%%%%%%%%%%%%%%%
%%% Acknowledgements. 
%%%%%%%%%%%%%%%%%%%%%%%%%%%%%%%%%%%%%%%%%%%%%%%%%%%%%%%
\Acknowledgements{We thank the anonymous referees for their constructive comments that highly improved the manuscript.
YL acknowledges the financial support by the Natural Science Foundation of China (Grant No. 11973090), and the science research grants from 
the China Manned Space Project with NO. CMS-CSST-2021-B06. GHMB and MF acknowledge funding from the European Research Council (ERC) under the 
European Union's Horizon 2020 research and innovation program (grant agreement No. 757957). GR acknowledges support from the Netherlands Organisation 
for Scientific Research (NWO, program number 016.Veni.192.233) and from an STFC Ernest Rutherford Fellowship (grant number ST/T003855/1).
We thank Tilman Birnstiel, Guo Chen, Ke Zhang and Richard Teague for insightful discussions. We acknowledge the DSHARP team for making the calibrated 
CASA measurement sets, fiducial images, and the scripts used for calibration and image cleaning, available for the public. ALMA is a partnership 
of ESO (representing its member states), NSF and NINS, together with NRC, MOST and ASIAA, and KASI, in cooperation with the Republic of Chile. 
The Joint ALMA Observatory is operated by ESO, AUI/NRAO and NAOJ.}

%%%%%%%%%%%%%%%%%%%%%%%%%%%%%%%%%%%%%%%%%%%%%%%%%%%%%%%
%%% Conflict of interest. 
%%%%%%%%%%%%%%%%%%%%%%%%%%%%%%%%%%%%%%%%%%%%%%%%%%%%%%%
\InterestConflict{The authors declare that they have no conflict of interest.}

%%%%%%%%%%%%%%%%%%%%%%%%%%%%%%%%%%%%%%%%%%%%%%%%%%%%%%%
%%% Supplements. 
%%%%%%%%%%%%%%%%%%%%%%%%%%%%%%%%%%%%%%%%%%%%%%%%%%%%%%%
%\Supplements{}

%%%%%%%%%%%%%%%%%%%%%%%%%%%%%%%%%%%%%%%%%%%%%%%%%%%%%%%
%%% Reference section.
%%% citation in the content using "some words~\cite{1,2}".
%%% ~ is needed to make the reference number is on the same line with the word before it.
%%%%%%%%%%%%%%%%%%%%%%%%%%%%%%%%%%%%%%%%%%%%%%%%%%%%%%%

\bibliographystyle{scichina}
\bibliography{hd163296}

\begin{appendix}

\renewcommand{\thesection}{Appendix}

\section{More information about the SED models}
\label{sec:moresed}

In Sect.~\ref{sec:sedmodel}, we modeled the SED of the HD\,163296 disk to constrain the maximum grain size ($a_{\rm max}$) for the LGP. Two models with 
$a_{\rm max}\,{=}\,1\,\rm{mm}$ (model \texttt{S1}) and $a_{\rm max}\,{=}\,1\,\rm{cm}$ (model \texttt{S2}) are shown in Figure~\ref{fig:bestsed}. 
Because the SED analysis is not the key of this study, we do not present all the information in the main section. 

\begin{figure}[H]
 \centering
 \includegraphics[width=0.48\textwidth]{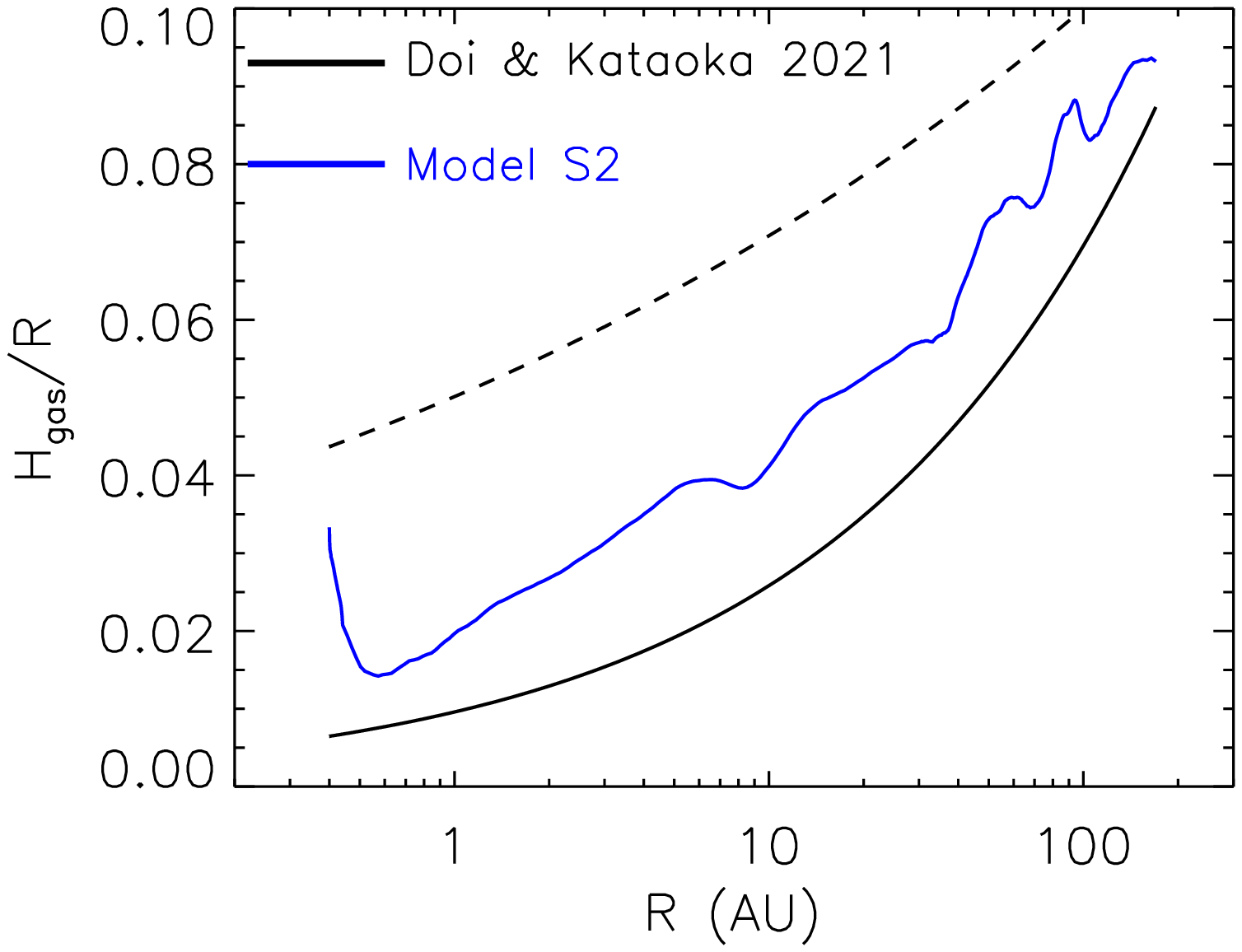}
 \caption{Gas scale height of the HD\,163296 disk. The blue line refers to the result of model \texttt{S2} (see Sect.~\ref{sec:sedmodel}), which is solved 
 under the condition of vertical hydrostatic equilibrium. The black line show the scale height used in Doi \& Kataoka \cite{Doi2021}. The dashed 
 curve, described as $H_{\rm gas}=10\,(R/100\,\rm{AU})^{1.15}$, is the typical disk geometry found from modeling the SEDs of T Tauri disks.}
\label{fig:s2hgas}
\end{figure}

\begin{figure}[H]
 \centering
 \includegraphics[width=0.48\textwidth]{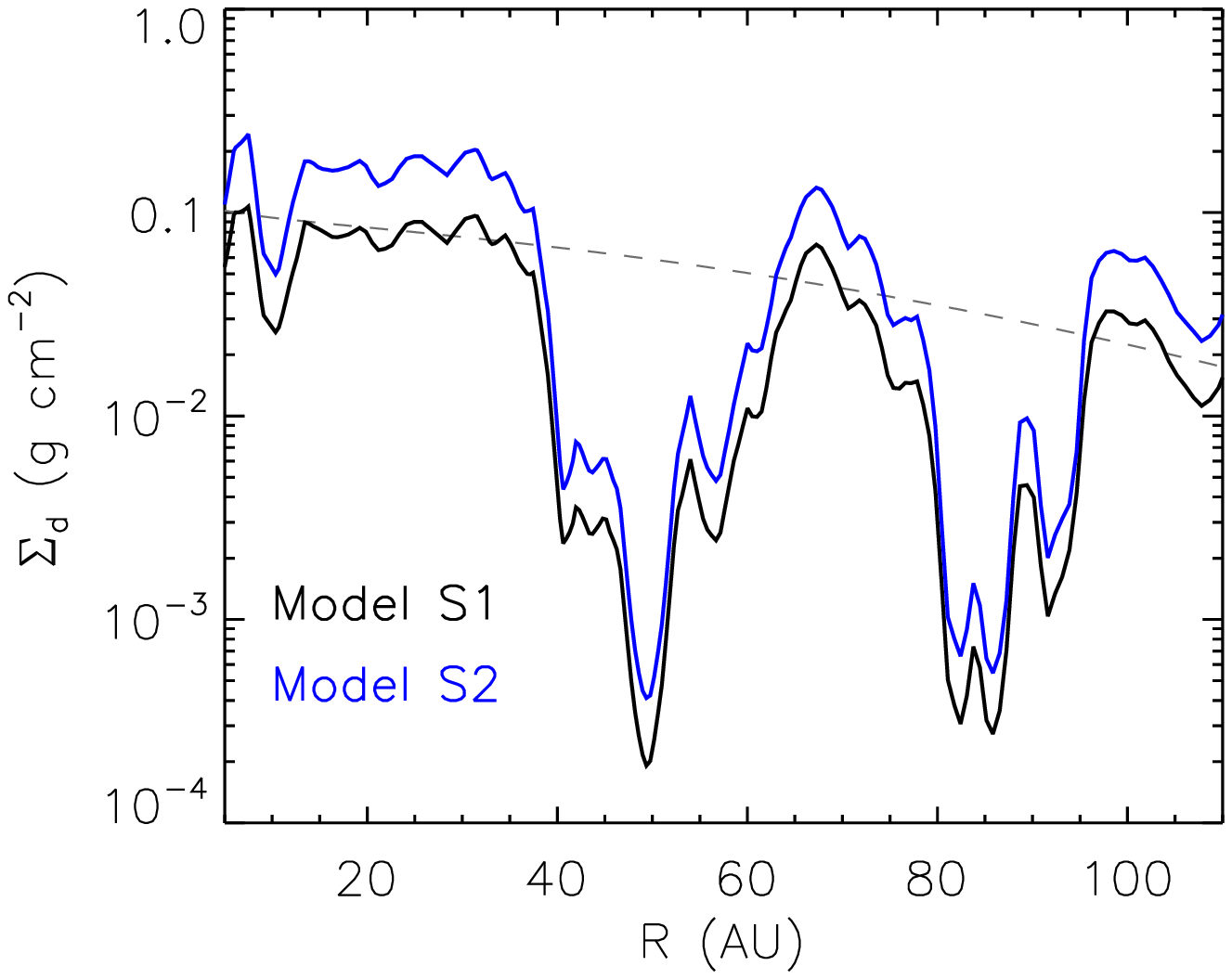}
 \caption{Dust surface densities of model \texttt{S1} (black solid line) and model \texttt{S2} (blue solid line). The grey dashed line represents the initial 
 density profile used in the iteration process, see Sect.~\ref{sec:surdens}.}
\label{fig:s1s2surdens}
\end{figure}

The blue solid line in Figure~\ref{fig:s2hgas} shows the gas scale height of model \texttt{S2}. The black solid line indicates the assumption made 
by Doi \& Kataoka \cite{Doi2021}, which is also our initial choice for $H_{\rm gas}$ in the iteration (see Sect.~\ref{sec:surdens}). The black dashed 
line stands for a typical profile found from modeling the SEDs of T Tauri disks. Figure~\ref{fig:s1s2surdens} shows the reconstructed surface 
densities ($\Sigma_{\rm d}$) for both models. As can be seen, they follow a similar pattern. However, the surface densities of model \texttt{S2} are 
systematically larger than those of model \texttt{S1}. This is because the mass absorption coefficient at a wavelength of 1.25\,mm for a dust grain 
population with $a_{\rm max}\,{=}\,1\,\rm{mm}$ is larger than that for a population of dust grains with $a_{\rm max}\,{=}\,1\,\rm{cm}$. 
Therefore, higher surface densities are required to fit the observed millimeter flux when $a_{\rm max}\,{=}\,1\,\rm{mm}$.

\end{appendix}

\end{multicols}
\end{document}